\documentclass[proceedings]{JHEP3} 
\PrHEP{ hep2001}

\usepackage{epsfig,multicol}			

\newbox\mybox
\newcommand\fverb{\setbox\mybox=\hbox\bgroup\verb}
\newcommand\fverbdo{\egroup\medskip\noindent\fbox{\unhbox\mybox}\ }
\newcommand\fverbit{\egroup\item[\fbox{\unhbox\mybox}]}

\title{ElectroWeak Symmetry Breaking in Tevatron Run II}

\author{\speaker{Giorgio Chiarelli for the CDF and D0 Collaborations
 }  

\\
	INFN,Sez.di Pisa, S.Piero a Grado(Pisa), Italy, Via Livornese 1291, 56010\\
	E-mail: \email{giorgio.chiarelli@pi.infn.it}}

\conference{FERMILAB-CONF-01/327-E}
\abstract{
The Tevatron Run II will provide CDF and D0 with a large dataset of
$p\overline{p}$ interactions 
collected at $\sqrt{s}=2 $TeV. We discuss the opportunity for the two experiments to
improve the understanding of electroweak and top physics in the first
years of data taking
(Run IIa, 2fb$^{-1}$) in view of the upgrades of the detectors.
We also discuss the prospectives
for a Higgs discovery at the Tevatron in view of the Run IIb data 
taking period which
will deliver an additional of about 13 fb$^{-1}$ to each experiment.
}

\begin{document} 
\section{Run II of the Tevatron Collider}
The successful Run I of the Tevatron in 1992-1996 led to the
upgrade of the accelerator complex. The new run, started in March 2001, is known as 
Run II and is expected to collect 2 fb$^{-1}$ in a first phase (Run IIa)
and, after 
another set of improvements to machine and detectors, gather additional 13
fb$^{-1}$ or so in what is known as Run IIb. 
The net results of the upgrades are an increase in energy (from $\sqrt 
s=$1.8 to 2 TeV) and in instantaneous luminosity (from
a typical 10$^{31}$ in Run I up to $5\times$10$^{32}$cm$^{-2}$s$^{-1}$
in Run IIb). The 
average number of interactions per crossing is kept low by decreasing the
inter bunch distance from 3.5 $\mu s$ of Run I down to 396 ns and
which will
eventually become 132 ns.

In order to match the technical challenges posed by those changes, as well as to 
exploit the physics capabilities of the Tevatron, both CDF and D0 underwent a series 
of upgrades~\cite{cdfup,doup}. Here we just mention a few of them, which are of 
special interest to the electroweak physics.
CDF completely redesigned its front-end electronics and DAQ system to
match the 132 ns 
interbunch. The trigger was also completely rebuilt, online tracking
for hight $P_t$ tracks was moved to the first level trigger, while a new 
special set of processors, 
providing the experiment with a second level trigger on tracks displaced
from the primary vertex (the 
Silicon Vertex Tracker) was constructed. Finally the whole tracking system
was 
completely rebuilt. It is now made of a new silicon tracker (7 layers
providing space points for $|\eta |<$2 at $2 \leq R \leq 28 $cm) and a
new central drift chamber with a stronger 3D reconstruction capability. 
Just to mention a few figure of merits, this system will allow 
tracking up to  $|\eta |<$2 (therefore doubling the Run I coverage), and
will 
increase the b tagging efficiency in top events to 65 \% per jet.
The D0 upgrade was even more substantial as a solenoid, providing a 2 T 
magnetic field, was added. Futhermore the tracking system was rebuilt with
a silicon vertex detector and a fiber tracker to fully exploit this new
situation (figure~\ref{d0upgr}). The D0 trigger, front end 
and DAQ systems were also redesigned to cope with the decreased bunch
spacing. In the end
the two experiments will have similar performances in terms of
hermeticity and tracking capabilities.
\FIGURE[r]{\epsfig{file=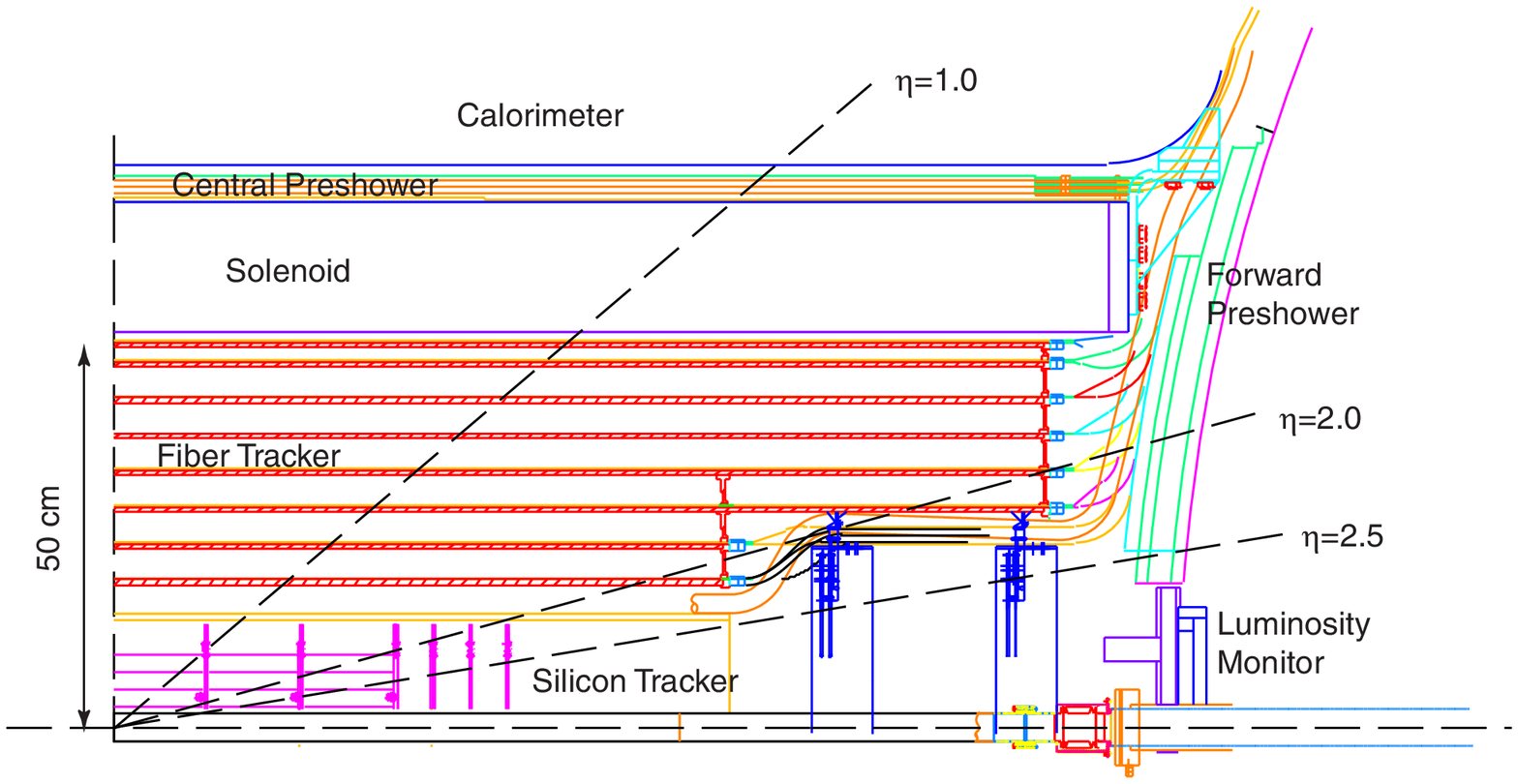,width=6.cm}
        \caption{D0 upgrade}
        \label{d0upgr}} 
To put things in perspective, compared to Run I the overall acceptance for
W,Z events
will be doubled and, for events containing a vector boson pair,
it will be tripled. 
\section{Gauge Boson Physics}
Run IIa with its 2 fb$^{-1}$ of data, will provide each experiment with
over 3$\times 10^6$ W's (e and $\mu$ channel), therefore allowing a
precise determination of many EW parameters.
One of the most important results will be a better
determination of the W mass. In Run I CDF measured $M_W$ with an
accuracy of 79 MeV, whereas D0's final accuracy was 84 MeV, bringing the
total uncertainty for this measurement (Tevatron only) to $\simeq$ 60 MeV,
with statistical and systematics on equal footing. 
However, since most of the systematic uncertainty depends upon the size of
control samples, we will greatly benefit from the significantly larger 
data set.
As an example we show (table~\ref{werror}) the contributions, essentially
all scaling with 
luminosity, for the CDF W$\rightarrow \mu \nu$ Run I
measurement~\cite{cdfwm}. 
\TABULAR[l]{|l|c|}{
        \hline \tt{\bf Source}&{\it {\bf uncert.}}\\
                  &{\bf {\it (MeV/c$^2$)}}\\
        \hline  \tt Fit statistics   & {\it 100}\\
        \hline  \tt Momentum scale  & {\it 85}\\
        \hline  \tt Recoil model& {\it 35}\\
        \hline  \tt Background&{\it 25}\\
        \hline  \tt Mom.resolution&{\it 20}\\
        \hline  \tt Selection bias&{\it 18}\\
        \hline}{\label{werror}{Run I $M_W$ uncertainties}
}
The biggest 
component (momentum scale) is obtained by fitting all the $l^+l^-$ 
invariant mass spectrum, while W asymmetry is used to constrain PDF and
the Z $P_t$ spectrum is the input to the W $P_t$ spectrum. With 2
fb$^{-1}$ we expect to reduce the overall uncertainty to
40 MeV/c$^2$, with 30 MeV/c$^2$ as a possible target.

The study of W and Z couplings will benefit from the wider acceptances and
the new 
features of the two detectors. The anomalous couplings in the ZV
(V=Z,$\gamma$) vertex are  parameterized in terms of $h_{30}$ and
$h_{40}$. Any value different from zero of those couplings would signal
new physics. In Run I D0 set a 95 \% C.L.limit for $h_{40}$ at about 0.3
and in Run IIa we expect to reach a limit to better than 0.03 by using
several hundreds of 
reconstructed Z$\gamma$ $\rightarrow  ee (\mu \mu) \gamma$ events and 
a few $ZZ\rightarrow  4 l$  detected.
In the W sector we expect to collect about 100 events WW $\rightarrow
l\nu l\nu$ and 
about 30 events WZ $\rightarrow lll \nu$. Scaling from Run I analysis, 
we expect to 
set limit for anomalous couplings, parameterized as $\Delta k$ and
$\lambda$ to $-0.12 < \Delta k
< 0.18$ for $\lambda=$0 and $|\lambda|<0.09$ for $\Delta k =0$.
One of the most important checks of the SM still to be performed, is the
detection of the {\it zero amplitude} in the W$\gamma \rightarrow \gamma
l \nu$ process. Due to specific SM cancellations, a dip appears in the
angular distribution of the final states ($l$,$\gamma$). CDF saw 
hints of this effect in Run I and we expect to collect about 3000
$W\gamma \rightarrow l\nu \gamma$ in Run IIa, therefore being able to
isolate this process.

\section{Top Physics in Run II}
Right after the discovery of the top quark, both experiments swiftly moved on to 
measure its properties (cross section, mass and couplings). Some of those 
measurements were already systematic-limited at the end of Run I. We
expect most of those systematics to scale with statistics as 
they are determined by the size of control samples and therefore
studies of the 
sixth quark will greatly benefit from the increase of luminosity and in
c.o.m. energy. Raising the energy from $\sqrt{s}=1.8$ to $2$ TeV, the
production cross section increases by 40\%, as the $gg$ 
scattering acquires a more prominent role. 
\TABULAR{|l|r|}{
        \hline \tt{\bf Channel}&{\it \bf CDF/D0}\\
        \hline  \tt Dilepton(e,$\mu$)& {\it 155}\\
        \hline  \tt Dilepton($\tau$)& {\it 19}\\
        \hline  \tt lepton$+\geq$ 3jets& {\it 1520}\\
        \hline  \tt lepton$+\geq$ 4 jets&{\it 1200}\\
        \hline  \tt lepton$+\geq$ 3jets+1btag&{\it 990}\\
        \hline  \tt lepton$+\geq$ 3jets+2btag&{240}\\
        \hline}{\label{topy} Yield of $t\overline{t}$ events in
2fb$^{-1}$} 
Table~\ref{topy} shows 
the number of top events expected in 2 fb$^{-1}$ per experiment. To put
things in perspective, CDF collected 
9 candidates dilepton events in Run I, and 76 were used in the l+4j
sample (the most important one to determine $M_{top}$) to 
reconstruct the top mass. The larger statistics will be obtained thanks to a wider
acceptance and better tracking capability. The improved b-tagging efficiency 
(for example at
D0 $\epsilon_b$ will be about 60 \% for $b$-jets with $P_t>$40 GeV/c) will 
greatly improve, besides the determination of $M_{top}$, the
measurement of the cross
section and the study of the $Wtb$ vertex.
\TABULAR{|l|r|}{
        \hline \tt{\bf Source}&{\it \bf GeV/c$^2$}\\
        \hline  \tt Jet En. scale& {\it 4.4(2.2)}\\
        \hline  \tt ISR and FSR& {\it 1.8(1.0)}\\
        \hline  \tt background& {\it 1.3(0.5)}\\
        \hline  \tt b-tag bias&{\it 0.4}\\
        \hline  \tt PDF&{\it 0.3}\\
        \hline  \tt Total&{4.9(2.5)}\\
        \hline}{\label{topmerr}CDF determination of $M_{top}$, systematics
in 2 fb$^{-1}$.
In parenthesis Run I results.}        
The measurement of $M_{top}$ is
dominated by systematics, with the biggest one being the jet energy scale. The
best way to tackle this problem is to identify a signal on which calibrate the energy scale.
In Run I CDF was able to reconstruct a $W \rightarrow jj$ peak in its
tiny $t\overline{t}$ double $b$-tagged sample and to detect the
$Z\rightarrow
b\overline{b}$ signal in its dijet sample. Both experiments plan to do
even better in Run II by using a dedicated trigger to select 
$Z\rightarrow b\overline{b}$ events. In this way the two vector bosons
will be used to set the jet energy scale and to check $b$-specific
corrections. Given that,
we expect the uncertainty to be reduced by a factor 2 
(see table~\ref{topmerr} for CDF estimates of expected systematics in Run
IIa; D0's number are similar). Together with the precise determination
of $M_W$
this will set stringent
limit on the Higgs mass (see figure~\ref{mwmtop}).

Another topic in top physics which will be explored in Run II is the precision 
determination of the top cross section, which is currently a $\simeq 25 
\%$ measurement and which will be determined to better than 10\% at the
end of Run IIa. The comparison between collider data and Monte Carlo will
improve
with the availability of larger samples of $b$ tagged events, the
wider acceptance and unprecedented MC samples both in level of detail and 
size.
Studies of the $Wtb$ vertex in Run I were performed at CDF. 
Both top helicity and ratio of branching fractions 
$R=\frac{t\rightarrow Wb}{t\rightarrow Wq}$ 
$=\frac{|V_{tb}|^2}{|V_{tb}|^2+|V_{td}|^2+|V_{ts}|^2}$
were measured~\cite{vtb,doh}. CDF 
\FIGURE{\epsfig{file=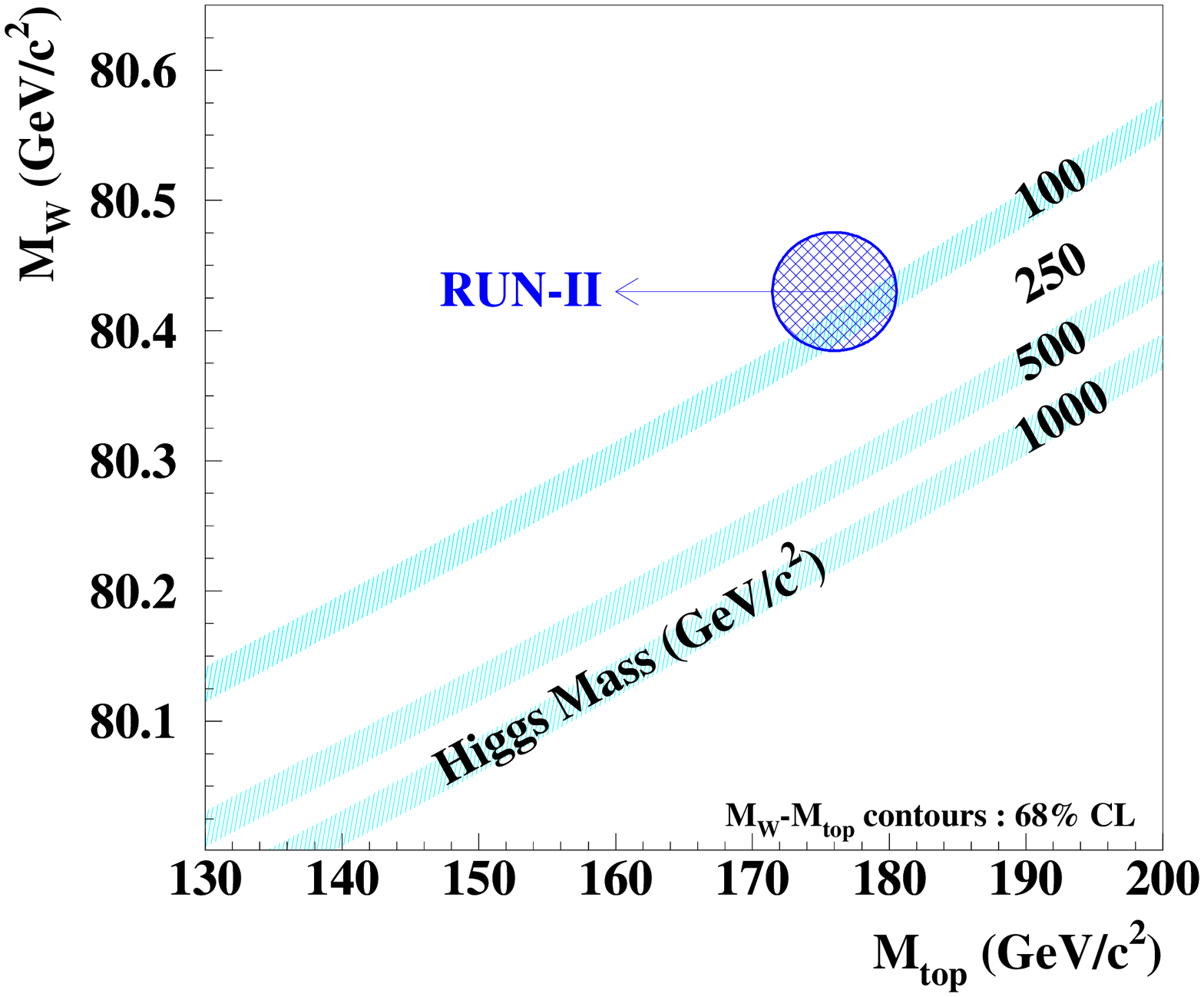,width=7.cm}
        \caption{SM Higgs as a function of $M_W$ and $M_{top}$. Run
IIa (2 fb$^{-1}$), one experiment only.}
        \label{mwmtop}}  

measurement of $R$ to 26 \% allowed an
indirect determination of the CKM matrix element $|V_{tb}|$ to $\simeq 15$\%.
In Run II we expect to reduce the uncertainty on $R$ to 6 \% with a corresponding
uncertainty on $|V_{tb}|$ to about 3\%. A  
direct determination of $|V_{tb}|$ (15\% accuracy) can be obtained by 
identification of the single top production process, which however stands as an elusive
process. The understading of forward tracking and tagging will play a decisive role
in this search.  

\section{Higgs Searches in Run II}
Over the last two years a lot of effort was dedicated to better understand
the chances 
to detect a SM Higgs particle at the Tevatron. Emphasis was set on the 
low mass ($<130$ GeV/c$^2$) region, where the Higgs decays almost
completely into 
$b\overline{b}$ pairs. 
Although (figure~\ref{higgsxsec}) the dominant production
process is the direct production 
through $gg$ fusion, the large QCD background renders this channel
unfeasible. 
\FIGURE[l]{\epsfig{file=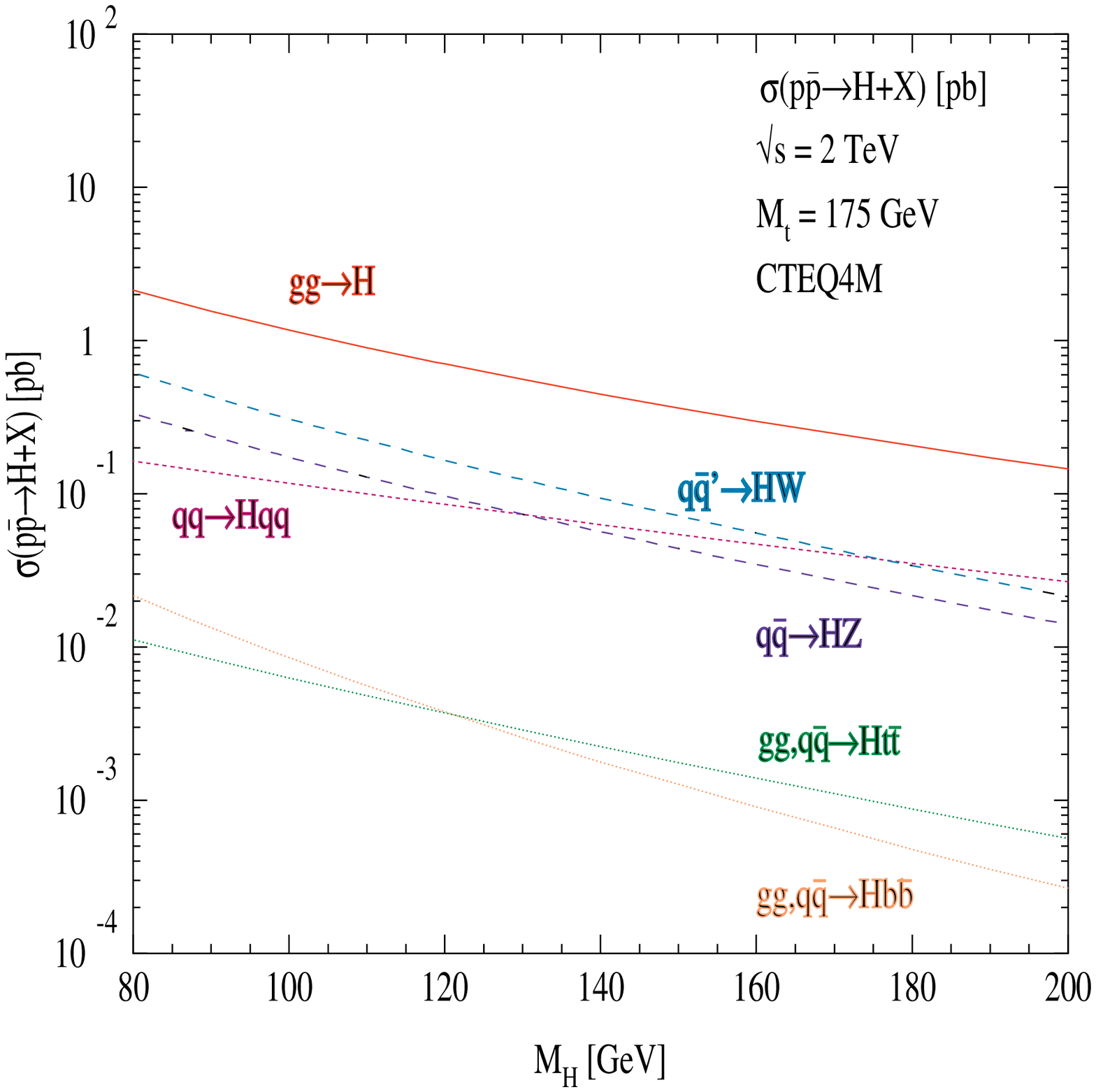,width=5.6cm}
        \caption{SM Higgs production}
        \label{higgsxsec}}    
The associate production $VH$
($V=W,Z$), where a leptonic decay of the $V$ 
would provide a clean trigger and bring to a favourable S/B ratio, is more 
promising.
A detailed study, based on a
mix of Monte Carlo paremeterization, Run I data analysis and comparison
with 
simulation of the CDF and D0 detector, was performed in the last three
years. It is not possible here to describe all the studies
performed\cite{carena},
the summary is provided in tables 4 and 5
where the event yield in 15 fb$^{-1}$ is shown for the
WH and ZH case ($b\overline{b}$ background included). It is important to
keep in mind that {\it all}, but
the single top,
background processes were already seen or measured in
Run I. Therefore the calculation of the signal includes all known
effects. 
\DOUBLETABLE{\begin{tabular}{|l|r|r|r|}
        \hline \tt {\bf $M_H$ (GeV/c$^2$)}&{\it \bf 110}&{\it \bf 120}&{\it \bf 130 }\\
        \hline  \tt Signal& {\it 75}&{\it 60}&{\it 45}\\
        \hline  \tt $Wb\overline{b}$& {\it 435}&{\it 375}&{\it 285}\\
        \hline  \tt WZ& {\it 90}&{\it 60}&{\it 30}\\
        \hline  \tt $t\overline{t}$&{\it 225}&{\it 300}&{\it 330}\\
        \hline  \tt single top&{\it 105}&{\it 135}&{\it 135}\\
        \hline  \tt S/$\sqrt{B}$&{2.6}&{2.0}&{1.6}
\end{tabular}}
{\begin{tabular}{|l|r|r|r|}
        \hline \tt {\bf $M_H$ (GeV/c$^2$)}&{\it \bf 110}&{\it \bf
120}&{\it \bf 130}\\
        \hline  \tt Signal& {\it 69}&{\it 46}&{\it 31}\\
        \hline  \tt $Zb\overline{b}$& {\it 84}&{\it 69}&{\it 52.}\\
        \hline  \tt $Wb\overline{b}$& {\it 100}&{\it 81}&{\it 63}\\
        \hline  \tt ZZ&{\it 43.5}&{\it 3}&{\it 0.0}\\
        \hline  \tt $t\overline{t}$&{\it 70.5}&{\it 64.5}&{\it 52.5}\\
        \hline  \tt single top&{\it 79.5}&{\it 70.5}&{\it 57.}\\
        \hline  \tt S/$\sqrt{B}$&{2.4}&{2.0}&{1.5}
\end{tabular}}{Higgs yield in $WH$ channel}{Higgs yield in $ZH$ channel}
{\label{lowwh}}{\label{lowzh}}
The studies of the "high mass" region focused on the $gg \rightarrow
H\rightarrow W^*W^*\rightarrow l^+l^-\nu\nu$ channel where the price paid
for the low cross   
section is offset by the very small background. An additional step was to add the 
associate production of H and W or Z: $p\overline{p} \rightarrow
W(Z)H\rightarrow 
l^{\pm}l^{\pm}jjX$. By vetoing on $b$-tagged jets (produced in 
$t\overline{t}$ events), the 
backgrounds are reduced to a manageable level.
The final outcome of these studies is shown in figure~\ref{Higgscomb}. The 
95 \% C.L. limit as well as the 3$\sigma$ evidence and 5$\sigma$ discovery
curves
are shown as a function of the Higgs mass and of the integrated
luminosity. The two experiments have been combined. There is a window of
opportunity in the low and intermediate
Higgs mass region, if the Tevatron can provide an integrated
luminosity in excess of 10-15 fb$^{-1}$ before the start of the LHC. 
In order to do so 
the Beams Division of Fermilab initiated an 
intensive R\&D program to demonstrate the 
feasibility of delivering about 4 
fb$^{-1}$ per year. 
At the same time both CDF and D0 
\vskip .1cm
\FIGURE{\epsfig{file=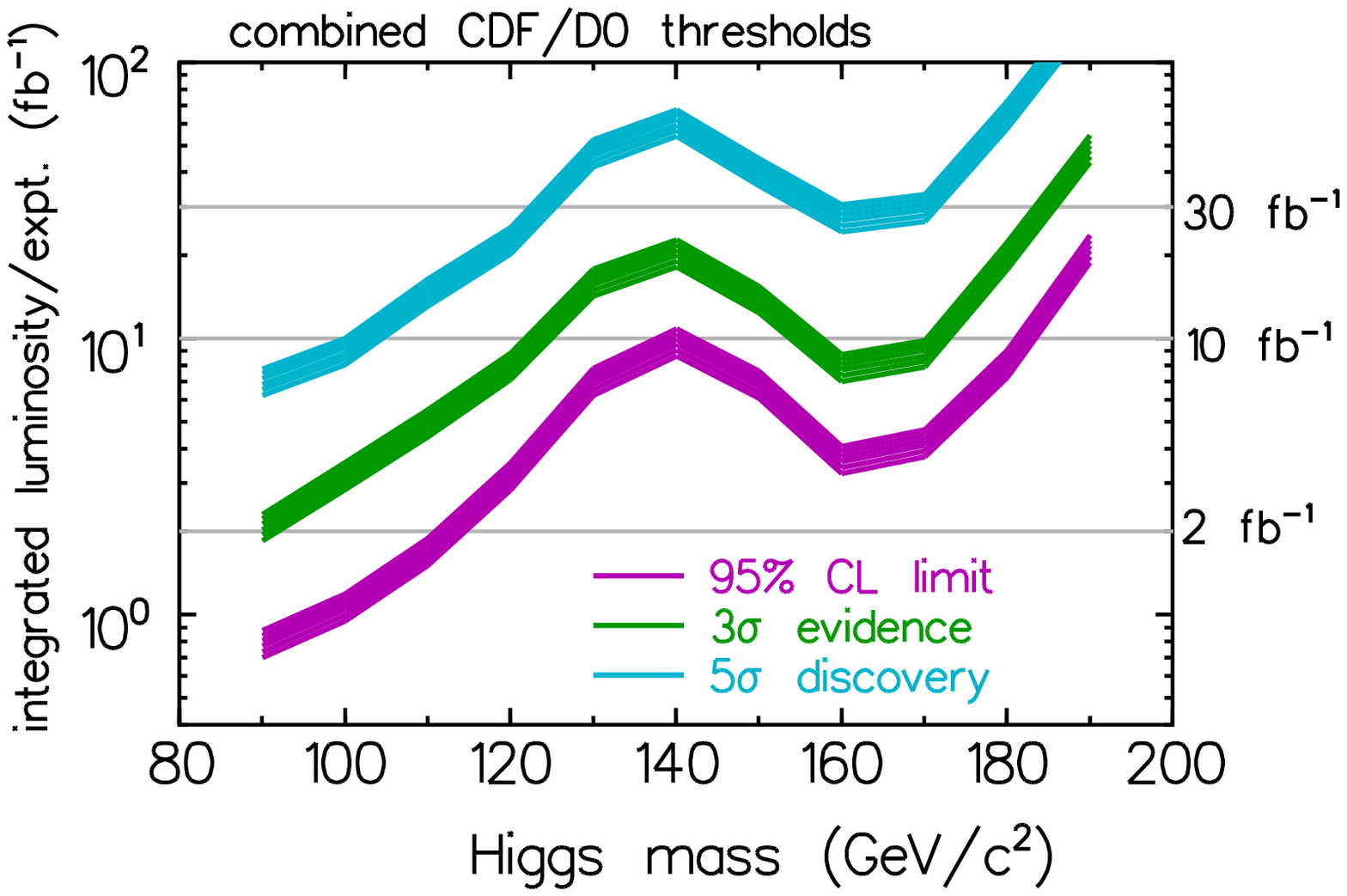,width=8.cm}
        \caption{SM Higgs expectations at the Tevatron}
        \label{Higgscomb}}   
started an intensive period of 
studies to understand which parts of their detectors need to 
be replaced 
to survive luminosities more than $7$ times larger than originally
planned. 
\section{Conclusion}
In the next years of Run II we expect to
gather more than 2 fb$^{-1}$ per experiment. 
Thanks to the very large
statistics precise determinations of the masses of the top quark and of
the $W$ vector boson will be  possible. This will allow to set stringent
limit
on the Higgs mass. Furthermore, as the Tevatron collider will be upgraded
to deliver $\simeq 4$ fb$^{-1}$ per year, there is a chance to isolate the
Higgs particle if its mass is low enough.


\end{document}